\documentclass[10pt]{iopart}
\usepackage[final]{graphics}
\usepackage{iopams}
\usepackage{amssymb}
\usepackage{amsfonts}
\usepackage{epsfig}
\usepackage{graphicx}
\usepackage{color}
\pdfoutput=1
\usepackage{subfigure}

\begin{document}
\title[Modelling of epitaxial film growth.]
{Modelling of epitaxial film growth with a 
Ehrlich-Schwoebel barrier dependent on the step height}
\author{F. F. Leal$^{1,2}$, S. C. Ferreira$^1$\footnote[1]{On leave at Departament de 
F\'{\i}sica i Enginyeria Nuclear, Universitat Polit\`ecnica de Catalunya,
Spain.},  and  S. O. Ferreira$^1$}
\address{$^1$Departamento de F\'{\i}sica - 
Universidade Federal de Vi\c{c}osa, 36571-000, Vi\c{c}osa,
Minas Gerais, Brazil}
\address{$^2$Instituto Federal de Ci\^encia, Educa\c{c}\~ao e Tecnologia - 
Campus Itaperuna, Rodovia BR 356 - KM 03, 28300-000, Itaperuna, 
Rio de Janeiro, Brazil.}

\ead{silviojr@ufv.br}

\begin{abstract}
The formation of mounded surfaces in epitaxial growth is attributed to the
presence of barriers against interlayer diffusion in the terrace edges, known as
Ehrlich-Schwoebel (ES) barriers. We investigate a model for epitaxial growth
using a ES barrier explicitly dependent on the step height. Our model has an
intrinsic topological step barrier even in the absence of an explicit ES
barrier. We show that mounded morphologies can be obtained even for a small
barrier while a self-affine growth, consistent with the Villain-Lai-Das Sarma
equation, is observed in absence of an explicit step barrier. The mounded
surfaces are described by a super-roughness dynamical scaling characterized by
locally smooth (faceted) surfaces and a global roughness exponent $\alpha>1$.
The thin film limit is featured by surfaces with self-assembled
three-dimensional structures having an aspect ratio (height/width) that may
increase or decrease with temperature depending on the strength of step barrier.
\end{abstract}


\submitto{\it J. Phys.: Condens. Matter}


\section{Introduction}

\label{sec:intro}

The most of the experiments of film production is performed in conditions far
from equilibrium, in which kinetic phenomena usually overcome any thermodynamic
effects in the film properties. Morphology, one of the most important features in
the film production, is ruled by distinct diffusion mechanisms
\cite{krugbook,Evans2006}, deposition rate \cite{Hamouda2008} and temperature
\cite{Ferreira2006.APL,EslaniPRL2006} among other factors. In particular, the
emergence of self-assembled and/or three-dimensional structures in form of
mounds has been observed during the growth of  a wide diversity of films, ranging
from metals \cite{Jorritsma,Caspersen,Yon2010} to inorganic
\cite{Johnson,EslaniPRL2006} and organic semiconductor materials
\cite{ZorbaPRB2006,Hlawacek2008}. This mound instability may be an undesirable
feature in the production of atomically flat interfaces but becomes a valuable
tool to produce structured surfaces, both systems with important technological
applications.

The three-dimensional growth is attributed to kinetic barriers that promote the
unbalance between up and downhill currents in stepped surfaces. A number of
different mechanisms has been proposed to explain these barriers, but there are
only few cases in which the observation of an unstable growth can clearly
be associated to a particular barrier \cite{EslaniPRL2006}. In two seminal
papers, Ehrlich \textit{et al.} \cite{ehrlich:1966} and Schwoebel \textit{et
al.} \cite{schwoebel:1966} independently reported that adatoms lying in
the edge of terraces diffuse backwardly in the same layer more frequently than move to a
lower step. This experiment led to a widely accepted hypothesis of an additional
activation energy to downhill diffusion known as Ehrlich-Schwoebel (ES) barrier.
Besides the ES barrier, alternative mechanisms have been proposed to explain
mound instability. Examples include the short-range attraction of adatoms
towards ascending steps \cite{Amar1996} and fast edge diffusion
\cite{Murty2003,Chatraphorn,Pierre-Louis}. 
 
Kinetic Monte Carlo (KMC) is a standard simulation method used to investigate
morphology and dynamics during the epitaxial film growth \cite{Evans2006}. In a
KMC simulation, the events (deposition, diffusion, evaporation, etc.) are
implemented, one at a time, with the specific rates.  In particular, diffusion
is a thermally activated process occurring at a rate given by an Arrhenius law
$D=\nu_0 \exp(-E_D/k_B T)$, where $\nu_0$ is an attempt frequency, $T$ the
growth temperature and $E_D$ the diffusion activation energy
\cite{vvdenskyPRB1995}. An ES barrier is usually included in KMC simulations by
raising $E_D$ when the particles are lying in descending steps. This additional
energy effectively inhibits the downhill diffusion and, consequently, increases
the chance of nucleation of new layers in top of the terraces. The result is an
instability responsible by the emergence of three-dimensional structures
\cite{Villain}.

Molecular dynamic simulations show that an efficient upward diffusion can also
be present in epitaxial growth \cite{UpwardDif}. Additionally, molecular dynamic
simulations also show that the ES barrier can change substantially from monolayer
to multilayer steps \cite{liuAPL2002}. With these motivations, we investigate a
model with a step-height dependent ES barrier using KMC simulations. We are
interested in the effect of these barriers in both down and uphill diffusion in
multilayer steps.  We have organized the paper as follows. In section
\ref{sec:model}, we develop the model and present the simulation procedures. In
section \ref{sec:result}, the simulation results are presented and discussed.
Finally, our concluding remarks are done in section \ref{sec:conclu}.   

\section{Model}
\label{sec:model}

The KMC simulations were performed as follows. Particles are deposited on a
substrate kept at a constant temperature $T$. The substrate, represented by a
triangular lattice with periodic boundary conditions, is initially flat.
Deposition occurs normally to the substrate at a constant rate $F$ and obeys the
solid-on-solid restriction to prevent overhangs. The height profile $h_j$ is
therefore the number of particles adsorbed in the site $j$. The deposition
involves two steps: firstly a site is picked up at random and secondly a new
particle is deposited in the most energetically favourable (largest bond number)
site among the chosen one and its nearest neighbours (NN).  If there are
multiple options, one is randomly chosen. This deposition rule is equivalent to
the classical Wolf-Villain model \cite{WV}.

\begin{figure}[t]
\centering
 \includegraphics[width=9cm]{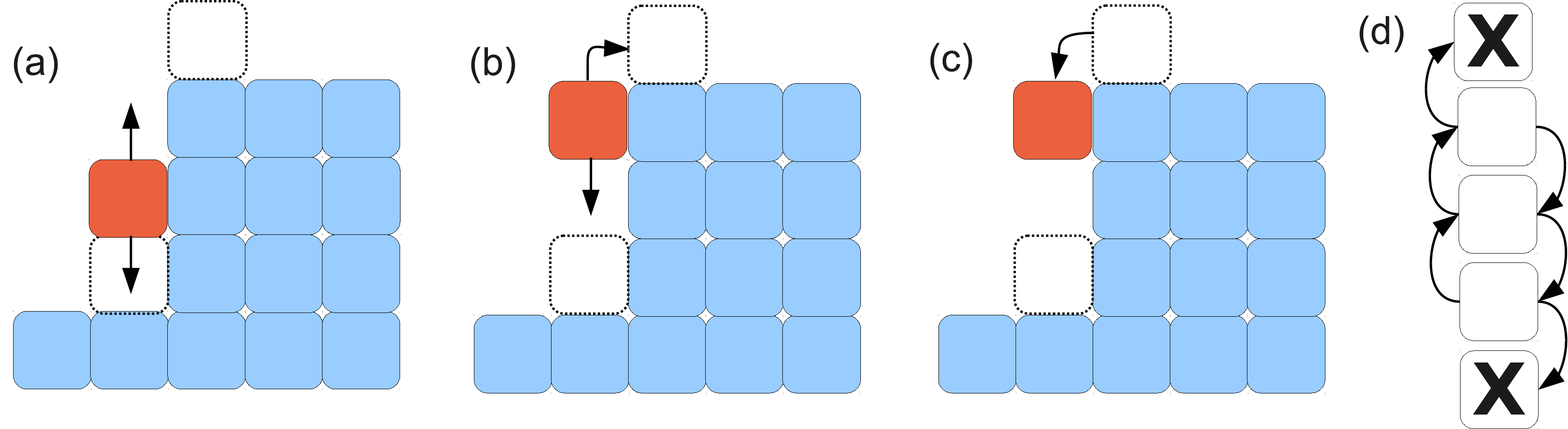}
\caption{The barrier model in multilayer steps. (a) A
particle in an ascending step detaches from its initial position (empty square)
and starts an unbiased random walk along the normal direction (red). (b)
The particle can be reflected in the top  due the presence of a ES barrier. (c)
A particle in a descending step (empty square) overcomes the ES barrier with
probability  $p=\exp(-E_b/k_bT)$ and then diffuses exactly as in (b).
(d) Random walk problem including the
absorbing/reflecting boundaries indicated by \textbf{X}.}
\label{fig:model}
\end{figure}

Intralayer diffusion from a site $j$ to $j'$, $\Delta h(j,j')= h_{j'}+1-h_j=0$,
has an activation energy $E_D(j,j')=E_0+n_jE_N$, where $E_0$ represents the
interaction with the substrate and $E_N$ the contribution of each one of the $n_j$
lateral bonds. The diffusion rate is therefore given by the Arrhenius law:
\begin{equation}
\label{eq:diff}
D(j,j') = \nu_0 \exp\left[-\frac{E_D(j,j')}{k_B T}\right].
\end{equation} 
Apart of the step barrier, this diffusion rule is similar to that used in the
model proposed by \v{S}milauer  and Vvedenski \cite{vvdenskyPRB1995}. Interlayer
diffusion is allowed for particles lying in both ascending or descending steps.
For monosteps, $\Delta h(j,j')=\pm 1$, the barrier is implemented as usually: an
additional barrier $E_b$ is present in a descending but not in an ascending
step. For steps higher than a monolayer, the interlayer diffusion is implemented
in two parts. If the diffusing particle lies in an ascending step, it firstly
detaches from its initial layer at a rate given by Eq. (\ref{eq:diff}) and then
executes a non-directed one-dimensional random walk, perpendicularly to the
substrate, as schematically illustrated in Fig.~\ref{fig:model}(a). A step
barrier is also present in the top of the step [Fig.~\ref{fig:model}(b)]. The
rule for a particle in a descending step is very similar except that it has to
overcome the ES barrier before attaching to the lateral
[Fig.~\ref{fig:model}(c)]. Notice that the barrier in the top makes the model
symmetric in relation to down and upward diffusion in multilayer steps, an
improvement of the usual rule without height restrictions. 

The determination of the probability of the particle moving to a neighbour site
or being reflected to its initial position involves the solution of a
one-dimensional random walk with absorbing/reflecting boundaries. When the
particle visits the boundaries, it becomes trapped with a given probability or is
reflected with complementary probability. This problem is schematically
illustrated in Fig. \ref{fig:model}(d) for a step of height $\ell =4$. In our
model, each boundary represents either the initial or neighbour sites. The
reflection is due to the ES barrier and, therefore, we set adsorption probabilities $1$ and
$p=\exp(-E_b/k_BT)$ for the the bottom ($i=0$) and the top ($i=\ell+1$) of the step,
respectively. Let $q(\ell+1|i)$ be the probability of the particle be trapped in
the top if it starts at a distance $i$ from the bottom. This random walk problem has an
exact solution explicitly developed in Ref.\cite{rwconfined}: 
\begin{equation}
 q(\ell+1|i)=\frac{i}{\ell+1/p}.
\end{equation}
For an ascending step [Fig. \ref{fig:model}(a)], it is easy to see that  the
probability of moving to the other layer is $q(\ell+1|1)$, while for
a descending step this probability becomes $q(\ell+1|\ell)\times p$. In latter
case, the factor $p$ takes into account the step barrier when the particle
crosses the kink [Fig.~\ref{fig:model}(c)]. The resulting probability
of diffusion through the step, for both cases, reads as:
\begin{equation}
\label{eq:phop}
P_{cross} =  \frac{p}{1+p(|\Delta h|-1)},~~~|\Delta h|\ge 2,
\end{equation}
where $\Delta h=h_{j'}+1-h_j$ is the height of the step to be crossed.

\section{Results}
\label{sec:result}

\begin{figure}[bt]
 \centering
 \includegraphics[width=12cm]{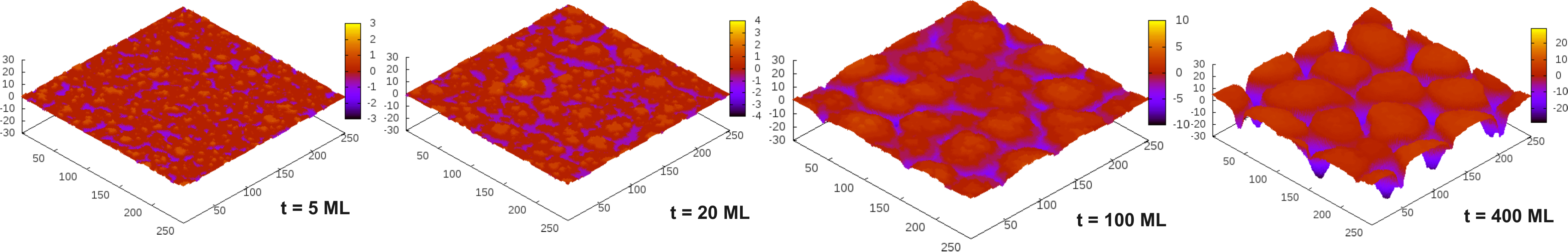}
 \caption{(Colour on-line) Surface morphologies for growth times varying from 5
to 400~ML. The growth temperature is $T = 573$~K and a mild step barrier
$E_b=0.02$~eV is used.}
 \label{fig:surface_time}
\end{figure}

The KMC simulations were performed with fixed parameters $E_0=1$~eV,
$E_{N}=0.11$~eV, $F=1$ monolayers (ML) per second, $\nu_0 = 10^{13}$~s$^{-1}$.
Temperatures $T=473,~523$ and $573$~K were simulated for deposition times of
$10^4$ ML. Except when explicitly mentioned, we report results for $256\times256$
sites. The initial surface evolution for a growth temperature of $T=573$~K and a
weak ES barrier of $E_b=0.02$~eV can be followed in Fig. \ref{fig:surface_time}.
One can see the surface coarsening and the emergence of three-dimensional
structures with a characteristic size in the form of mounds. For very long
times, however, the instability is highly enhanced turning the surfaces in
deeply grooved morphologies with large terraces.  

\begin{figure}[b]
 \centering
\subfigure[\label{fig:surf}]{
\begin{minipage}{8.5cm}
\includegraphics[width=8.5cm]{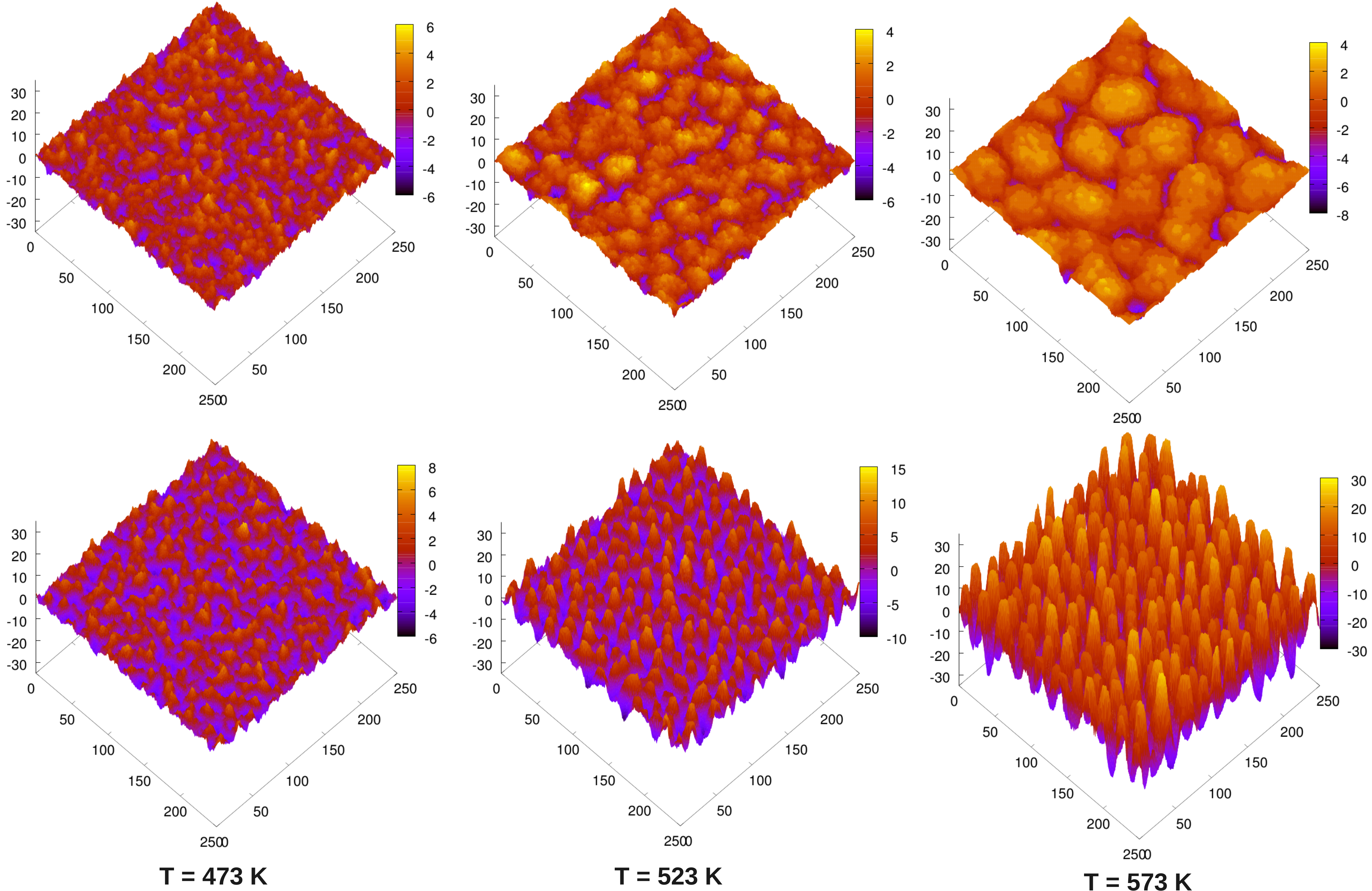}
\end{minipage}}~~
\subfigure[\label{fig:ilhas}]{
 \begin{minipage}{3cm}
 \includegraphics[width=3.0cm]{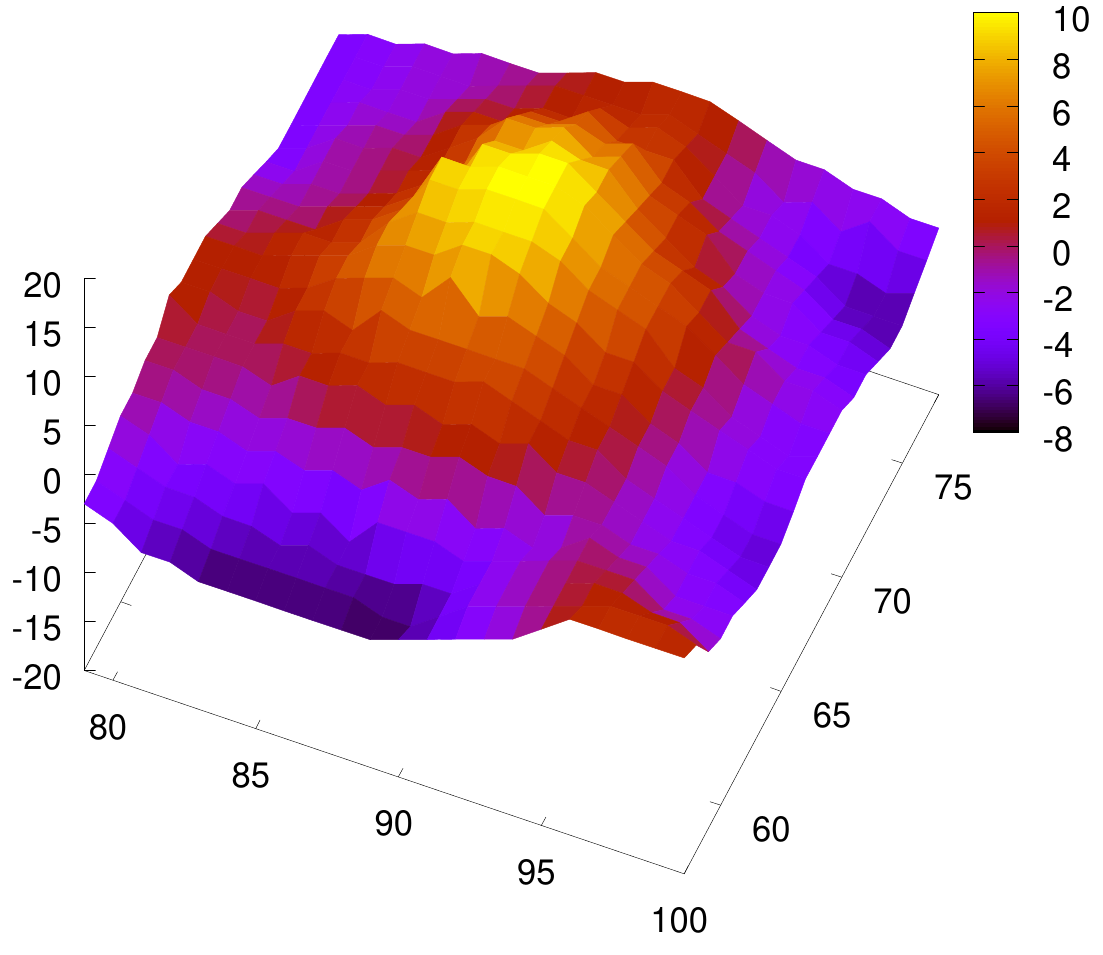} 
 \includegraphics[width=3.0cm]{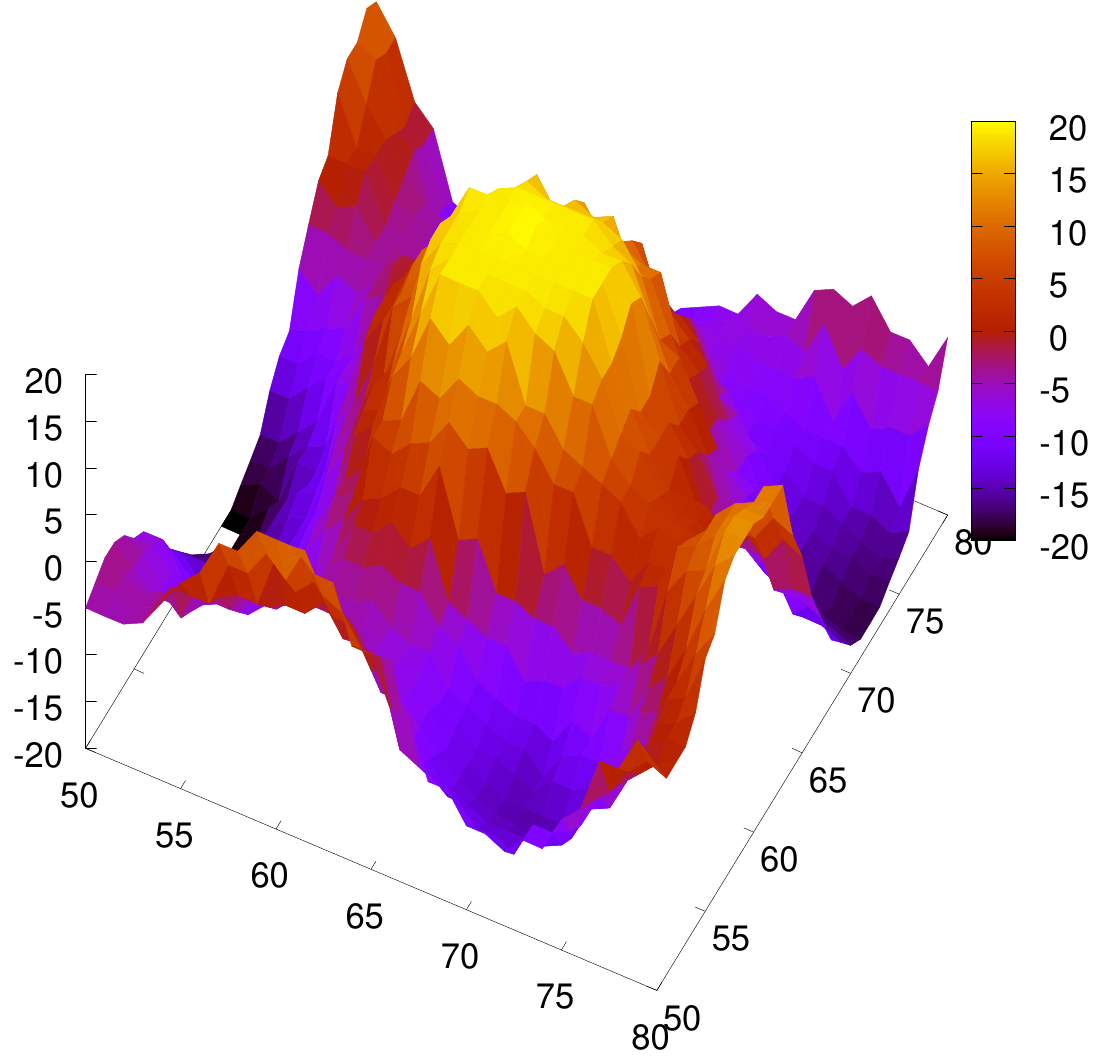}
 \end{minipage}}
 \caption{\label{fig:surface_temp} (Colour on-line) (a)~Surface morphology for
distinct growth temperatures after the deposition of 100~ML. The bottom and top
surfaces correspond to step barriers $E_b=0.02$~eV and $E_b=0.07$~eV,
respectively. (b)~Grain morphology for $E_b=0.07$~eV and growth temperatures
$T=523$~K (top) and $T=573$~K (bottom). Both grains are in the same scale.}
\end{figure}

\begin{figure}[bt]
 \centering
 \subfigure[\label{fig:wa}]{\includegraphics[width=6.0cm]{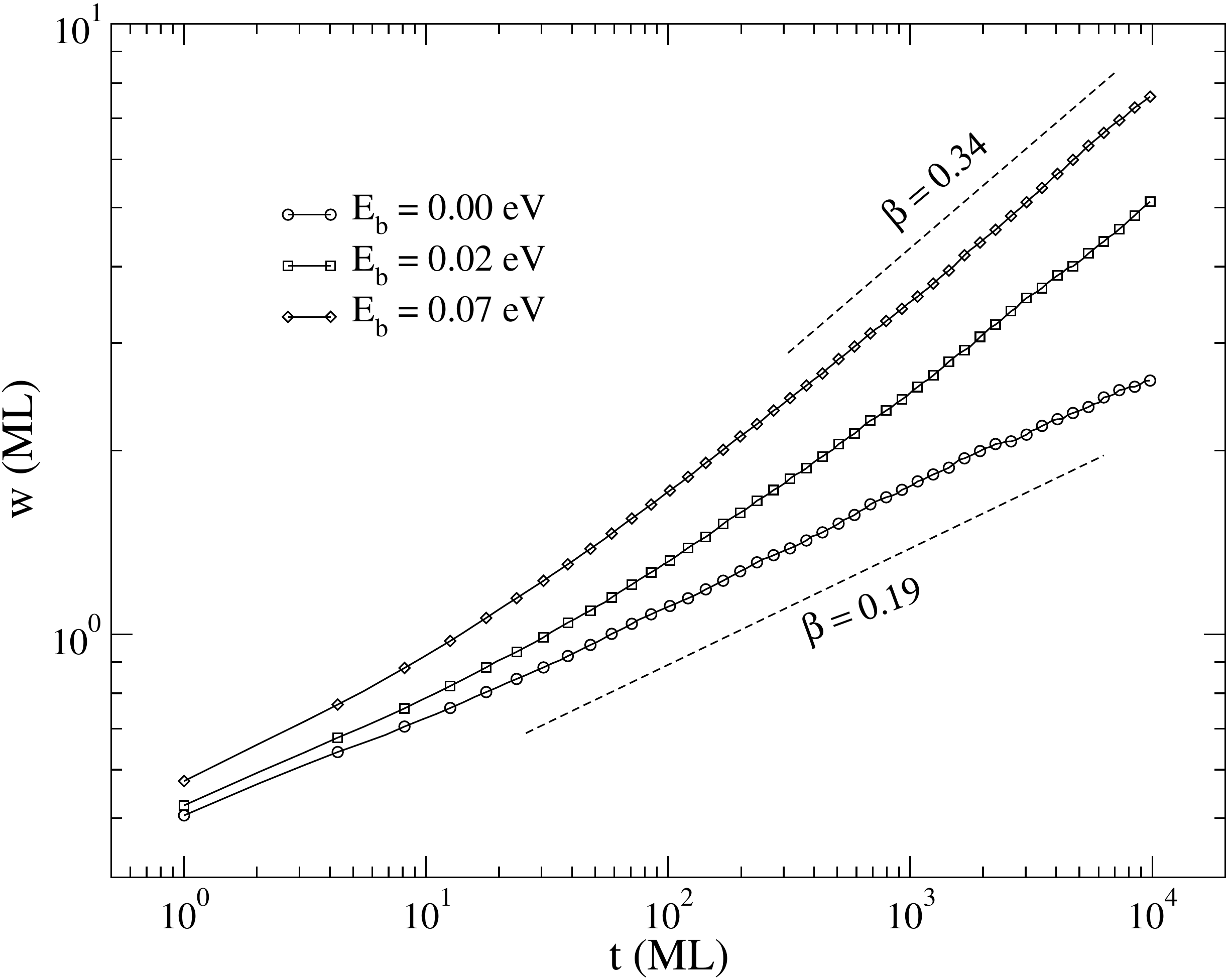}} 
 \subfigure[\label{fig:wb}]{\includegraphics[width=6.2cm]{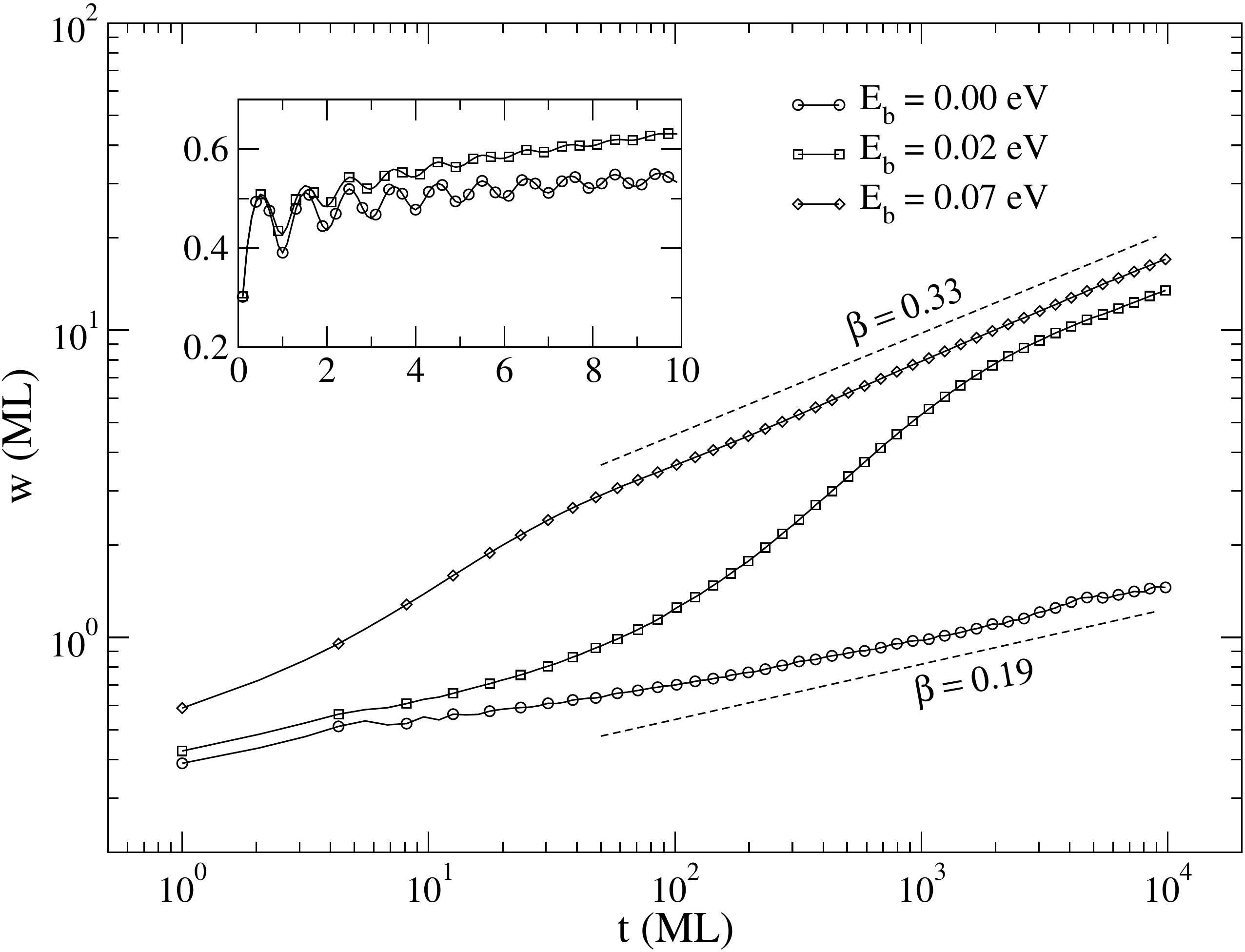}}
 \caption{\label{fig:w} Interface rms roughness evolution for (a) $T=473~K$ and
(b) $T=523~K$. The inset shows the initial layer-by-layer growth for $T=523$~K.
Dashed lines are scaling laws obtained from data regressions.}
 \end{figure}

A rich variety of spatio-temporal properties is observed in the thin film
phase. The present model has an intrinsic step barrier independent of the
parameter $E_b$ when $|\Delta h|\ge2$. If we set $E_b=0$, the system has a
transition from a rough (fractal) surface at low temperatures to an almost
layer-by-layer growth at high temperatures [inset of Fig. \ref{fig:wb}]. The
layer-by-layer regime is transient and becomes self-affine for asymptotic times,
as we will see later. However, a weak barrier of $E_b=0.02$~eV is sufficient to
induce a temperature-driven instability and to form mounds. In the top row of
Fig. \ref{fig:surf}, we can see the temperature effect for this weak barrier. The
mound aspect ratio (height/width) decreases with temperature, a behaviour also
observed in homoepitaxy of some semiconductor materials \cite{Ge_mound}. If a
step barrier $E_b=0.07$~eV is used instead, the surface morphology changes
considerably. The uphill diffusion becomes strong at higher temperatures
generating self-assembled three-dimensional structures, as shown in
the bottom of Fig. \ref{fig:surf}. Details are illustrated in Fig.
\ref{fig:ilhas}, where isolated grains for temperatures $T=523$ and 573~K are
shown. Higher temperatures make the grains higher resulting a large increase of
the aspect ratio.

\begin{figure}[bt]
 \centering
\subfigure[\label{fig:xi200}]{\includegraphics[width=6cm]{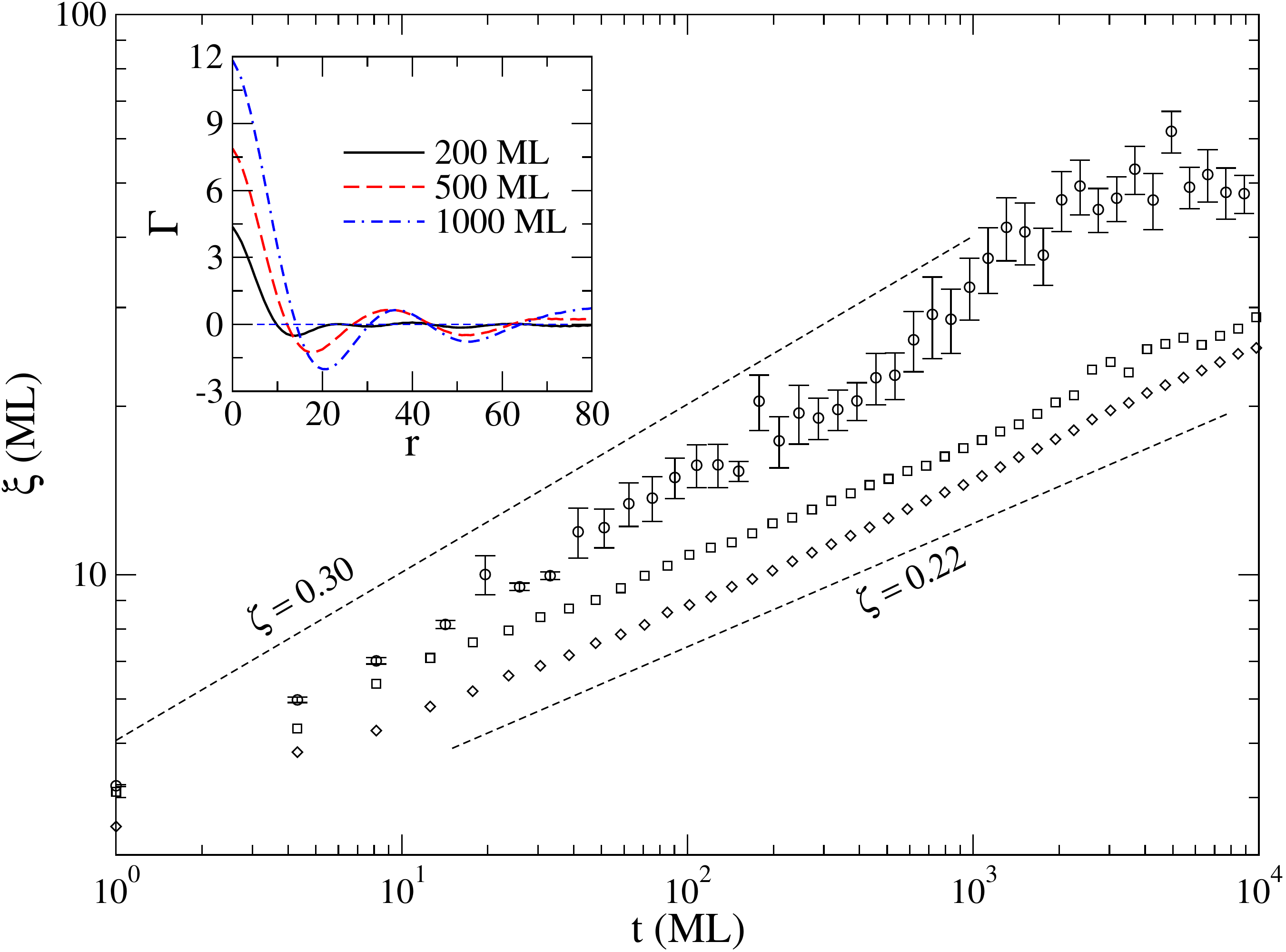}}
\subfigure[\label{fig:xi250}]{\includegraphics[width=6cm]{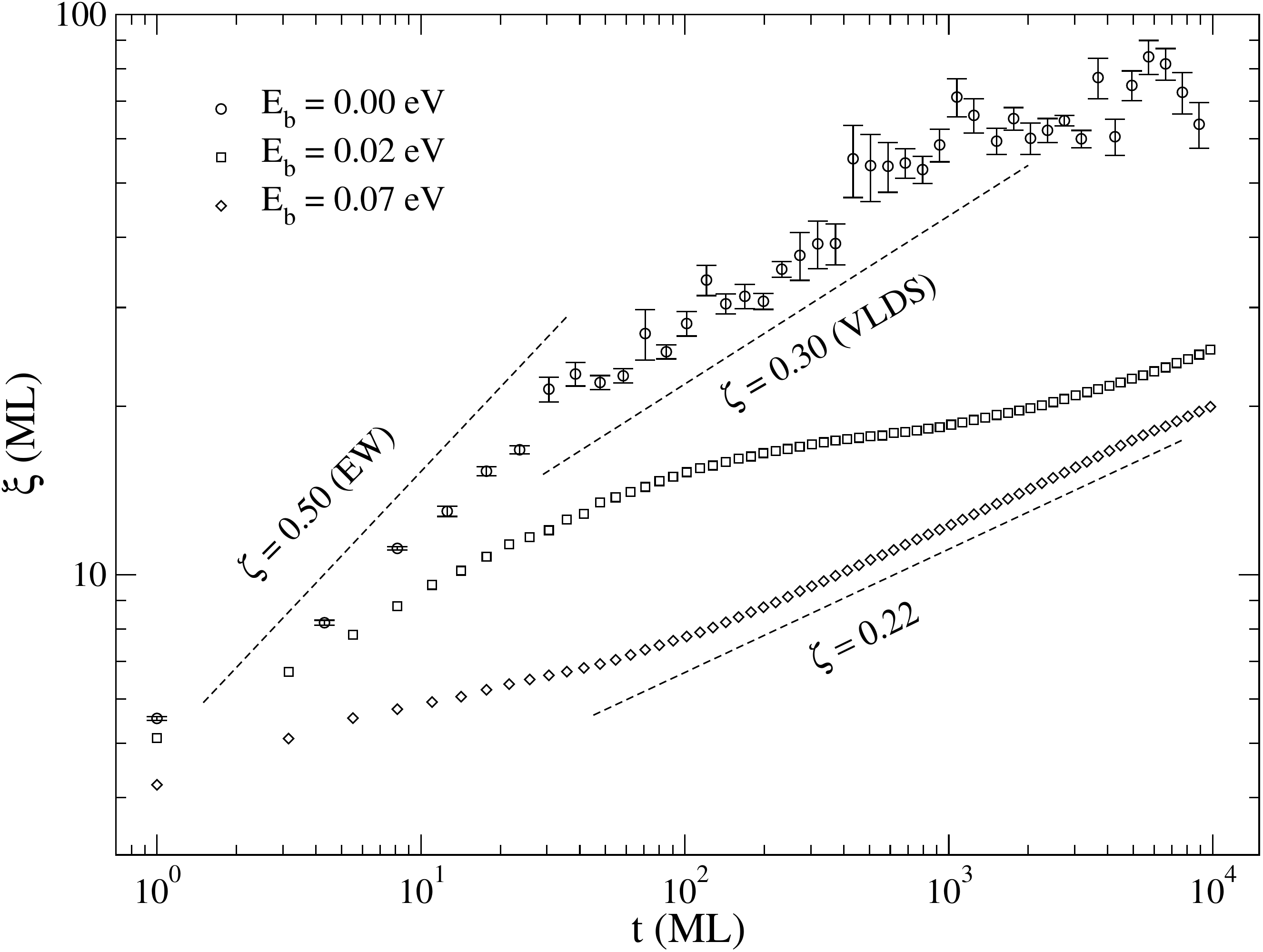}}
 \caption{\label{fig:correl} (Colour on-line) Characteristic length against time
for (a) $T=473$~K and (b) $T=523$~K using distinct step barriers. Inset shows
the correlation functions for different deposition times and $E_b=0.07$~eV. The
curves correspond to averages over 20 independent samples. The error bars
smaller than symbols were omitted. Dashed lines are power laws with the
indicated exponent as guides to the eyes.} 
 \end{figure}

The thick film limit also exhibits very interesting dynamics. A basic
quantity to characterize the surface morphology and dynamics in this regime is
the height-height correlation function defined as \cite{Murty2003}:
\begin{equation}
\label{eq:cor}
\Gamma(\mathbf{r}) = 
\langle h(\mathbf{x})h(\mathbf{x}+\mathbf{r})\rangle_\mathbf{x}, 
\end{equation}
where $h(\mathbf{x})$ is the surface height in the mean height referential and
$\langle\cdots\rangle_\mathbf{x}$ represents the average over the surface. The
first zero of $\Gamma(r)$, denoted by $\xi$, is a characteristic length  of the
surface. Typical correlation functions for distinct growth times are shown in the insertion to Fig.
\ref{fig:xi200}. The oscillating behaviour observed in the correlation functions
is the hallmark of mounded surfaces~\cite{krugbook,Murty2003}. The length $\xi$
is expected to increase as a power law of time, $\xi\sim
t^{\zeta}$, where $\zeta$ is the so-called coarsening exponent~\cite{Murty2003}.
Other basic quantity  to characterize the surface dynamics is the interface rms
roughness given by $w=\sqrt{\Gamma(0)}$. For a kinetic roughening, this quantity
grows as a power law of time, $w\sim t^\beta$, where $\beta$ is the growth
exponent. In dynamical scaling theory \cite{krugbook}, the coarsening and growth
exponents are related with the roughness exponent by the scaling relation
$\alpha=\beta/\zeta$.  These exponents are very useful because they may
determine the universality class of the system and, from the framework of the
generic dynamical scaling theory, they may connect scaling and morphological
properties of the surface~\cite{Ramasco}. 

For all investigated temperatures the scaling exponents change if an explicit
step barrier is present or not. For the lower temperature of $T=473~$K and in the
absence of an explicit step barrier ($E_b=0$), we found a  power law regime with
coarsening and growth exponents $\zeta = 0.32$ and $\beta = 0.19$,
respectively, as shown in Figs. \ref{fig:xi200} and \ref{fig:wa}. A
surface evolution for these parameters is shown in MOVIE 1 of the supplementary 
material. These exponent values are in good agreement with the exponents $\zeta=3/10$ and
$\beta=1/5$ obtained for the classical Villain-Lai-Das Sarma (VLDS) growth
equation \cite{Villain,LaiPRL91} 
\begin{equation} \frac{\partial h}{\partial t}
= -\nu\nabla^4h + \lambda\nabla(\nabla h)^2+\eta, 
\label{eq:VLDS} 
\end{equation}
where $\eta$ is a Gaussian noise. Similar exponents are obtained for higher
temperatures and $E_b = 0$. Interestingly, in the limit of zero temperature, our model
corresponds to the Wolf-Villain model \cite{Villain} that has a crossover
\cite{VvedenskyPRB2003,Alves2011}
to the universality class of the Edwards-Wilkinson (EW) equation
($\partial_t h = \nu \nabla^2 h+\eta$) with exponents $\beta=0$ (logarithmic
growth) and $\zeta=0.5$ in $d=2+1$~\cite{EW} .
Contrarily, our simulations at higher temperatures point out the opposite
behaviour: a small growth exponent (consistent with a logarithm growth)  and a
coarsening exponent close to $1/2$ are observed for early growth times whereas
$\beta\approx1/5$ and $\zeta\approx3/10$ are seen only at long times. The
results suggest that the diffusion, not the deposition rules, is governing  the
dynamics. 

\begin{figure}[t]
 \centering
 \includegraphics[width=5cm]{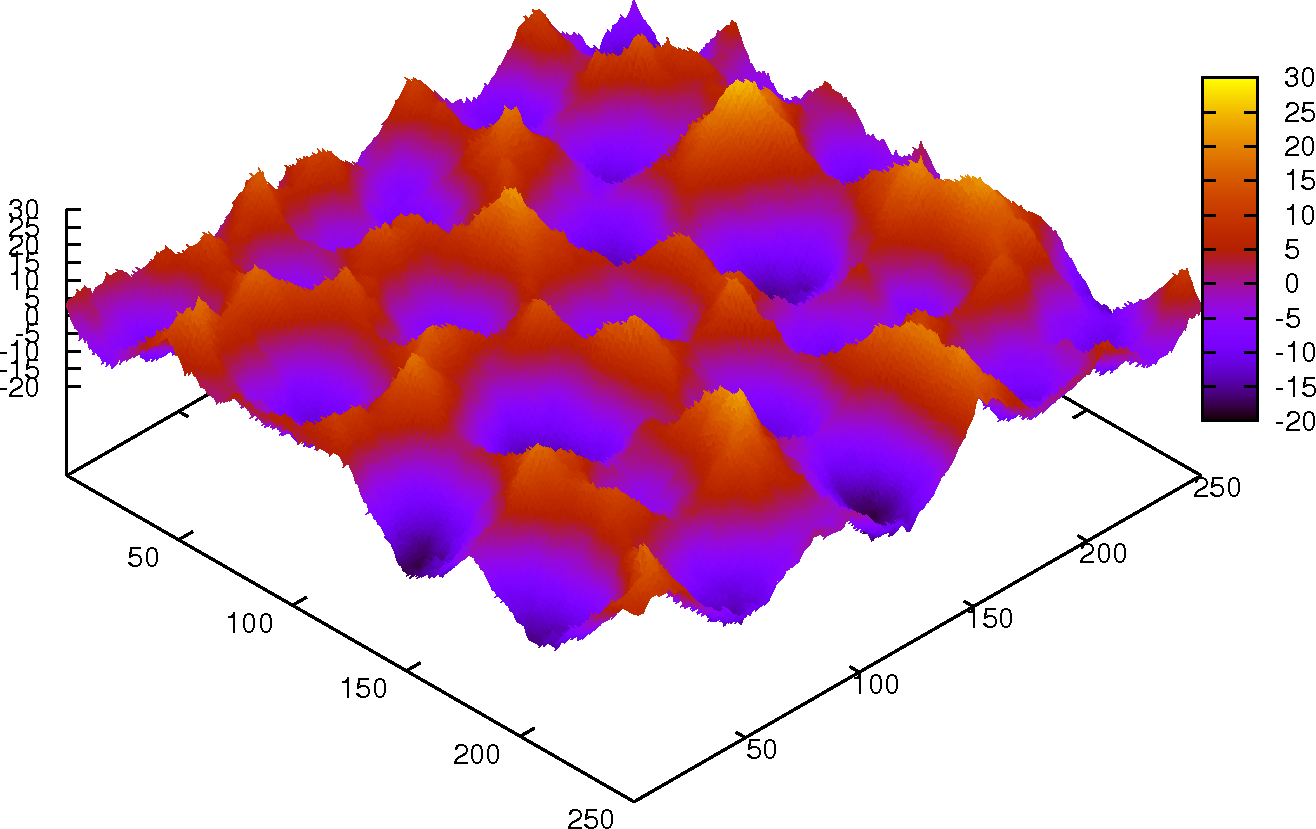} ~
 \includegraphics[width=5cm]{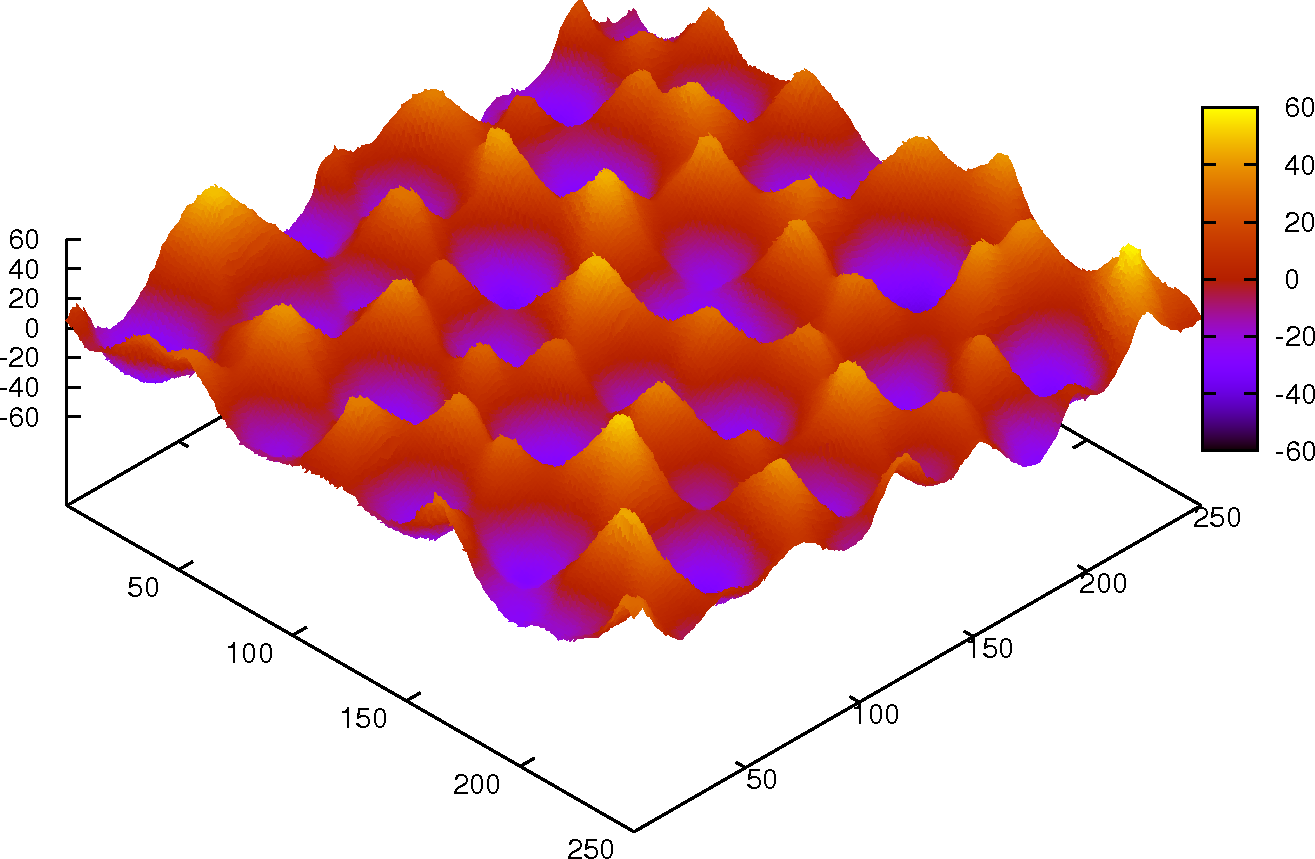}
  \caption{\label{fig:facet} (Colour on-line) Surface morphologies after 
   deposition of $10^4$ ML using an step barrier $E_b = 0.07$~eV at 
   temperatures $T=473$~K (left) and $T=523$~K (right).}
\end{figure}

In the presence of an explicit ES barrier, the scaling exponents  for $T=473$~K
and $T=523$~K converge (more slowly for $E_b=0.02$~eV than for 0.07~eV) to
$\beta = 0.33$ and $\zeta = 0.22$, as shown in Figs. \ref{fig:correl} and
\ref{fig:w}. A growth exponent $\beta=0.33$ were also obtained for the higher
investigated temperature of $T=573$~K, but the measured coarsening exponent,
$\zeta\approx 0.18$, seems not be the asymptotic value. Anyway, the corresponding
roughness exponents, given by $\alpha = \beta/\zeta$, are larger
than 1 implying that the explicit barrier causes the so-called
super-roughness dynamical regime \cite{Ramasco}. Super-roughness is featured by
a locally smooth surface with rms roughness in windows of size $\ell$ scaling as
$w(\ell)\sim \ell$ for $\ell<<\xi$. Left panel of Fig. \ref{fig:facet} shows
surfaces for a long deposition time obtained for $E_b=0.07$~eV. Similar
structures are obtained for a mild barrier $E_b=0.02$~eV, but the faceted
morphology is less evident. For $T=473$~K, the system evolves from a initially
rough  [Fig. \ref{fig:surf}] to a faceted morphology exhibiting pyramidal mounds
as can be seen in MOVIE 2 of the supplementary material. The
surface dynamics becomes richer for higher temperatures. Initially,
self-assemble and fast growing structures are formed due to the
instability caused by the step barrier. Thereafter, these structures start to
coalesce and to form large scale mounds also exhibiting pyramidal morphology. The
surface evolution can be followed in MOVIE 3 of  the supplementary material.

\section{Concluding remarks}
\label{sec:conclu}

The pattern formation in semiconductor films produced with epitaxial techniques
is a subject of basic and technological interest. The formation of
three-dimensional structures is related with the so called Ehrlich-Schwoebel
(ES) barrier. This barrier is usually implemented in kinetic Monte Carlo (KMC)
simulations by reducing the diffusion rate of particles in descending steps. In
the present work, we investigate the homoepitaxial growth of films performing KMC
simulations where the ES barrier depends explicitly on the step height. In
addition, our model includes a barrier to interlayer diffusion in ascending
steps when their heights are larger than a monolayer.  

The model yields surfaces with mounded morphology even for a mild step barrier
of $E_b=0.02$~eV, a value much smaller than the typical nearest neighbour
interactions of $E_N=0.11$~eV used in the present KMC simulations. However, a
layer-layer growth, lasting during the deposition of several layers, is obtained if
the explicit ES barrier is null. For an intermediary barrier $E_b=0.07$~eV,
self-assembled three-dimensional structures are obtained. The aspect ratio
(height/width) of these structures decreases with temperature for a mild barrier but
increases for the intermediary one. 

We have also investigated the thick film limit using the  growth and coarsening
exponents, defined as $w\sim t^\beta$ and $\xi\sim t^{\zeta}$, respectively. In
the absence of an explicit barrier, the interface undergoes kinetic roughening
with exponents $\beta\approx 0.19$ and $\zeta\approx0.30$ what is consistent with the
non-linear Villain-Lai-Das Sarma (VLDS) equation \cite{Villain,LaiPRL91} for
molecular beam epitaxy (MBE). In the presence of a step barrier, the exponent
changes to $\beta\approx0.33$ and $\zeta\approx 0.22$ what turns out a roughness
exponent $\alpha=\beta/\zeta>1$ and states the so-called super-roughness
scaling regime \cite{Ramasco} for the surface evolution. This regime is featured by
large pyramidal structures that are observed in MBE systems like Ge(100)
\cite{Ge_mound} and Cu(100) \cite{Ernst}. Even though a large number of models
has been proposed to reproduce specific systems  with pyramidal
morphology~\cite{Evans2006}, our model is quite simple, robust and does not
require any kind of parameter tuning.

The present model addresses two features that, to our knowledge, were 
neglected in the broad literature of KMC modelling of epitaxial growth: the
influence of the step height in the ES barrier and the explicit barrier when the
particle is executing an uphill diffusion. This last one is particularly
interesting since the rules for the movement through a kink becomes symmetric in
relation to up or downhill movements [Fig. \ref{fig:model}(b)]. Finally, the
specific form of the interlayer diffusion probability [Eq. (\ref{eq:phop})] is
not essential to the surface dynamics. We have also simulated, but not shown in
this paper, a model where Eq. (\ref{eq:phop}) is replaced by $P_{cross}=p$ for
$|\Delta h|\ge2$ and observed that all morphological properties and scaling
exponents are preserved.

\section*{Acknowledgements}

This work was partially supported by the Brazilian agencies CNPq and
FAPEMIG.  S.C.F. thanks the kind hospitality at the Departament de
F\'{\i}sica i Enginyeria Nuclear/UPC.

\providecommand{\newblock}{}

\end{document}


\title[Supplementary material]
{Supplementary material of  ``Modelling of epitaxial film growth with a Ehrlich-Schwoebel barrier dependent on the step height''}
\author{F.F. Leal$^{1,2}$, S. O. Ferreira$^1$ and
S. C. Ferreira$^1$}

\begin{abstract}
We described the animations of the surface evolution generated with kinetic Monte Carlo simulations. 
\end{abstract}

\begin{description}
 \item [MOVIE 2] \texttt{T473Eb00L256\_small.avi} - Surface evolution for a substrate temperature of $T=473$~K in the absence of an explicit step barrier. 
 
 \medskip

\item [MOVIE 2] \texttt{T473Eb07L256\_small.avi} - Surface evolution for a substrate temperature of $T=473$~K and a step barrier of $E_b = 0.07$~eV. 
 
 \medskip
 
  \item [MOVIE 3] \texttt{T523Eb07L256\_small.avi} - Surface evolution for a substrate temperature of $T=523$~K and a step barrier of $E_b = 0.07$~eV. Scale changes during the movie for sake of visibility.

 
  
\end{description}